\documentclass[prl,twocolumn,amsmath,amssymb,superscriptaddress, bibnotes]{revtex4-2}

\usepackage{graphicx}% Include figure files
\usepackage{dcolumn}% Align table columns on decimal point
\usepackage{amsmath,amssymb,mathtools,bm,mathrsfs,url,amsthm,textcomp,amsfonts,color,pxrubrica,slashed,braket,siunitx, comment}
\usepackage[per-mode=symbol]{siunitx}  % ← / をスラッシュ表記に
\DeclareSIUnit\bohrmagneton{\ensuremath{\mu_B}}
\usepackage[version=4]{mhchem}
\usepackage{tikz}
\usepackage{tikz-feynhand} 
\usepackage{comment}
\usepackage{indentfirst}
\begin{document}

\title{Superconducting Properties on Two-dimensional Quasicrystal (Ta$_{0.95}$Cu$_{0.05}$)$_{1.6}$Te Studied with $^{125}$Te-NMR}

\author{H.~Matsudaira}
\email{matsudaira.hiroyasu.34a@st.kyoto-u.ac.jp}
\author{S.~Kitagawa}
\author{K.~Ishida}
\email{ishida.kenji.5n@kyoto-u.ac.jp}
\affiliation{Department of Physics, Kyoto University, Kyoto 606-8502, Japan}
\author{Y.~Tokumoto}
\author{K.~Tomiyama}
\author{K.~Edagawa}
\affiliation{Institute of Industrial Science, University Tokyo, Tokyo 153-8505, Japan}

\date{\today}

\begin{abstract}
Physical properties in the normal and superconducting (SC) state are investigated with $^{125}$Te-nuclear magnetic resonance (NMR) measurements in a quasicrystal \ce{(Ta_{0.95}Cu_{0.05})_{1.6}Te}, which was a recently discovered superconductor with the SC transition temperature $T_{\mathrm{c}}$ = 0.94 K. 
The nuclear spin-lattice relaxation rate $1/T_1$ shows a coherence peak just below $T_{\mathrm{c}}$, followed by an exponential decrease down to 0.1 K.
The overall temperature dependence of $1/T_1$ is in good agreement with an $s$-wave SC model with a SC gap slightly smaller than the BCS value.
However, the coherence peak is unusually small, which may be attributable to a reduced Bogoliubov peak theoretically predicted for quasicrystals.
Furthermore, $^{125}$Te-NMR spectra show almost no broadening nor shift in the SC state,  suggesting that an unusual SC state such as parity mixing might be realized in the Ta$_{1.6}$Te superconductor.  
\end{abstract}

\maketitle
%\section{Introduction}
%SC:superconducting
Quasicrystals (QCs) are a unique class of solids exhibiting long‐range quasiperiodic order and crystallographically forbidden rotational symmetries, with lacking conventional translational periodicity. 
Since their discovery by Shechtman \textit{et al.} in 1984 in rapidly quenched \ce{Al-Mn} alloys \cite{Shechtman_PRL_1984}, QCs have renewed our understanding of solid‐state order and symmetry \cite{Levine_PRL_1984,Levine_PRB_1986,Socolar_PRB_1986,internationalcrystal_1992_report}.
%Recently, various intersting physical phenomena is observed in QC, such as Quantum criticality robust against pressure\cite{Deguchi_Natmat_2012}, or various magnetic order\cite{Tamura_JACS_2021,Tamura_naturephys_2025,Fisher_JAC_2000}.
The realization of the long‐range electronic ordered state \cite{Tamura_JACS_2021,Tamura_naturephys_2025,Fisher_JAC_2000, Deguchi_Natmat_2012}, especially superconductivity in QCs \cite{Kamiya_Natcomm_2018}, is one of the most fundamental and intriguing topics in condensed matter physics, because fundamental concepts of solid‐state physics have been developed on the basis of the translational symmetry. 

The first clear evidence of superconductivity in QC was reported by Kamiya \textit{et al.} \cite{Kamiya_Natcomm_2018}. 
They observed zero resistivity, Meissner signal, and a specific‐heat jump at superconducting (SC) transition temperature $T_{\mathrm{c}} =$ \SI{0.05}{K} in an icosahedral \ce{Al-Zn-Mg} QC. 
Nonetheless, the extremely low $T_{\mathrm{c}}$ prevented us from detailed investigations of the SC state in QCs. 
In addition, theoretical studies on the attractive Hubbard model on a Penrose lattice have predicted that superconductivity in QCs can host intrinsically inhomogeneous order parameters, spatially extended Cooper pairs, and unusual SC state \cite{Sakai_PRR_2019,Takemori_PRB_2020,Sakai_PRB_2017,Nagai_PRB_2022,Cao_PRL_2020}. 
%Moreover, it is predicted that a magnetic field can induce vortex pinning without impurities \cite{Nagai_PRB_2022} and Fulde–Ferrell–Larkin–Ovchinnikov (FFLO)–like states with nonzero Cooper‐pair momentum in QC superconductors \cite{Sakai_PRR_2019}.
To verify these predictions experimentally, it has been desired to synthesize a QC superconductor with a higher $T_{\mathrm{c}}$.

A major breakthrough occurred when superconductivity in the van der Waals (vdW)–layered \ce{Ta_{1.6}Te} dodecagonal QC was discovered \cite{Tokumoto_Natcommun_2024}. 
%\ce{Ta_{1.6}Te} is a layered material, the layers of which consist of \ce{Ta-Te} units arranged in a two‐dimensional dodecagonal tiling structure, separated by vdW gaps \cite{Conrad_Angew_1998,Cain_PNAS_2020}. 
This material was originally discovered by Conrad \textit{et al.} \cite{Conrad_Angew_1998}.
Tokumoto \textit{et al.} fabricated bulk polycrystalline samples via reaction sintering, and conclusively demonstrate bulk superconductivity in \ce{Ta_{1.6}Te} with $T_{\mathrm{c}} = \SI{0.98}{K}$ \cite{Tokumoto_Natcommun_2024}.
%—nearly twenty times higher than that of the \ce{Al-Zn-Mg} icosahedral QC (i-QC) \cite{Kamiya_Natcomm_2018}. 
Measurements of electrical resistivity, magnetic susceptibility, and specific heat on \ce{Ta_{1.6}Te} confirmed a weak‐coupling, $s$‐wave SC state \cite{Tokumoto_Natcommun_2024}.
%, with an electron–phonon coupling constant $\lambda_{\mathrm{ep}}\approx 0.52$ estimated from McMillan’s formula \cite{Tokumoto_Natcommun_2024}.
Soon after the discovery, Terashima \textit{et al.} revealed that the upper critical field $H_{\mathrm{c2}}(T)$ of \ce{Ta_{1.6}Te} QC decreases linearly over the entire temperature range below $T_{\mathrm{c}}$, with a large slope $\mathrm{d}H_{\mathrm{c2}}/\mathrm{d}T =$ \SI{-4.4}{T/K}, and $H_{\mathrm{c2}}(0) =$ \SI{4.7}{T}, which far exceeds the Pauli-depairing field $\sim$ \SI{1.8}{T} \cite{Terashima_npj_2024}. 
These findings raise intriguing questions about the impact of quasiperiodicity on superconductivity.
%not only demonstrate that \ce{Ta_{1.6}Te} is a rare example of a thermodynamically stable bulk QC superconductor, but also 
Thus, further experiments probing local electronic state in QC superconductors are necessary. 

Nuclear magnetic resonance (NMR) is a powerful probe to study the local electronic state. 
We carried out \ce{^{125}Te}–NMR on the dodecagonal quasicrystal \ce{(Ta_{0.95}Cu_{0.05})_{1.6}Te}, measuring both the Knight shift and the spin–lattice relaxation rate $1/T_1$ to investigate the SC state from the microscopic viewpoint. 
The exponential $1/T_1$ decay with a suppressed coherence peak below $T_c$, consistent with the $s$-wave superconductivity with a nodeless full gap, whose Bogoliubov peak is seemingly weakened by quasiperiodicity.
Furthermore, we observe almost no NMR spectrum shift and line broadening in the SC state, suggesting an unconventional SC state, which would be directly tied to the quasiperiodic structure.
These findings provide new microscopic insight into the SC properties realized in quasiperiodicity.

\begin{figure}
    \centering
    \includegraphics[width=\linewidth]{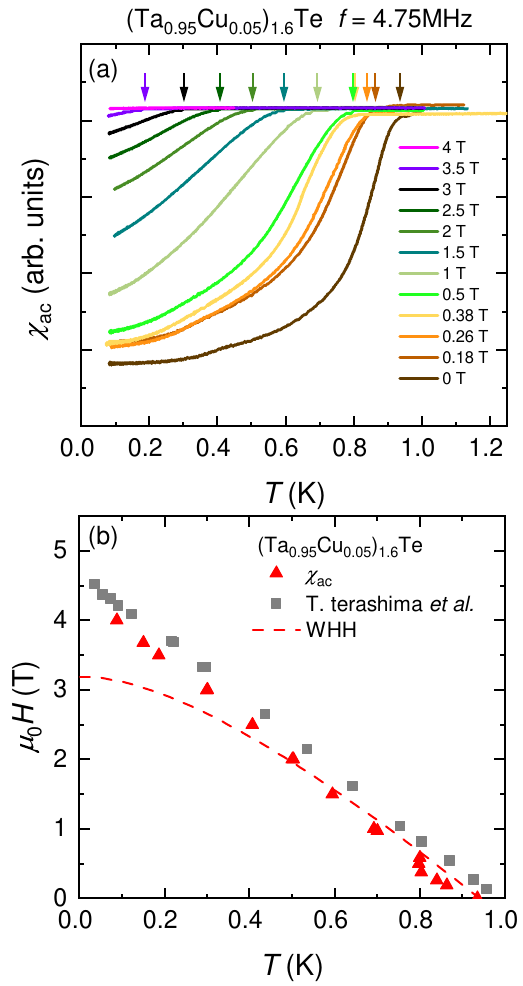}
    \caption{(a) Temperature dependence of the ac susceptibility under various fields. Arrows indicate the SC transition temperature $T_{\mathrm{c}}$ determined from the onset of the diamagnetic signal. (b) Magnetic-field–temperature $H$–$T$ phase diagram of \ce{(Ta_{0.95}Cu_{0.05})_{1.6}Te} together with the $H_{\mathrm{c2}}$ result of \ce{Ta_{1.6}Te} by Terashima \textit{et al.} \cite{Terashima_npj_2024}. The dashed line shows the orbital critical field based on WHH theory \cite{Werthamer_PR_1966}.}
    \label{meissner}
\end{figure}

%\section{Experimental}
We used a polycrystalline QC \ce{(Ta_{0.95}Cu_{0.05})_{1.6}Te} sample prepared by a reaction-sintering method. 
A small amount of \ce{Cu} was intentionally added to suppress the synthesis of impurity phases.
The sample was cooled using either a liquid \ce{^{4}He} cryostat or a \ce{^{3}He}/\ce{^{4}He} dilution refrigerator.
To characterize SC properties, we performed ac-susceptibility measurements by recording the resonance frequency of the NMR tank circuit with the sample. 
The resonance frequency $f$ is converted into $\chi_{\mathrm{ac}}$ by $\chi_{\mathrm{ac}} = 1 / f^2$.
The SC transition temperature $T_{\mathrm{c}}$ was determined by the onset of the diamagnetic signal, as shown in Fig.~\ref{meissner} (a).

$^{125}$Te (nuclear spin $I=1/2$, gyromagnetic ratio $^{125}\gamma/2\pi = $ \SI{13.454}{MHz/T} \cite{GCCarter1976})-NMR measurements were carried out, using a conventional spin-echo method. 
%, and natural abundance 7.0\%
The Knight shift was determined from the peak maximum of each spectrum.
The magnetic field was calibrated by observing the $^{63(65)}$Cu [$^{63(65)}\gamma/2\pi = 11.285(12.089)$ MHz/T] signal from the copper NMR coil with $K =$ \SI{0.2385}{\%} \cite{GCCarter1976}.
The spin-lattice relaxation rate $1/T_{1}$ was measured using the saturation-recovery method. 
The details of how to determine $1/T_1$ are explained in supplemental material \cite{SM}.

%\section{Results and Discussions}
%%%%%%%%%%%%%%%%%%%%%%%%%%%%%%%%%%%%%%%%%%%%%%%%%%%
%%%%%%%%%%%%%%%%%%%%%%%%%%%%%%%%%%%%%%%%%%%%%%%%%
Figure \ref{meissner} (a) shows temperature ($T$) dependence of the ac susceptibility measured under various magnetic fields ($H$).
The clear SC diamagnetic signal was observed below $T_c$ up to 3.5 T.
The $H$–$T$ phase diagram, obtained from the ac susceptibility measurement, is displayed in Fig.~\ref{meissner}(b). 
The upper critical field $H_{\mathrm{c2}}(T)$ decreases almost linearly with slope $dH_{\mathrm{c2}}/dT = \SI{-4.9}{T/K}$, in good agreement with the previous result by Terashima {\it et al} \cite{Terashima_npj_2024}.
%As discussed by Terashima \textit{et al.}, such linearity of $H_{\mathrm{c2}}(T)$ may be attributed to spatial electronic inhomogeneities on the order of SC coherence length \cite{Carter_Solidstate_1981}or multi-band effect \cite{Gurevich_PhysicaC_2007}. 
This linearity might suggest the realization of unconventional SC state, such as Fulde–Ferrell–Larkin–Ovchinnikov (FFLO) state under high magnetic fields near $H_{c2}$, which was suggested by the theoretical work \cite{Sakai_PRR_2019}.
The extrapolated $\mu_{0}H_{\mathrm{c2}}(0)$ is \SI{4.6}{T}, corresponding to a coherence length of $\xi = 84.6$ \AA.
This suppressed coherence length is attributed to a short mean free path \cite{tinkham2004introduction}, which might be intrinsic to QCs. 
In QCs, structural inhomogeneities originating from quasiperiodicity may induce strong electron scattering and make mean free path shorter \cite{Klein1_EPL_990}.
%in fact the resistivity of \ce{Ta_{1.6}Te} is similar to dirty metal \cite{Tokumoto_Natcommun_2024,Terashima_npj_2024}. 
%This is quite consistent with previous results determined with the resistivity measurements \cite{Tokumoto_Natcommun_2024,Terashima_npj_2024}.
The zero-field transition temperature, $T_{\mathrm{c}}(0) = \SI{0.94}{K}$, is slightly lower than the reported values (\SI{0.98}{K}) \cite{Terashima_npj_2024}, reflecting the small amount \ce{Cu} added in our sample, but no secondary SC transition was observed, as shown in Fig.~\ref{meissner} (a).
Although the observed result is consistent with the Werthamer–Helfand–Hohenberg (WHH) theoretical curve in the small field region \cite{Werthamer_PR_1966}, clear deviation was observed as in the cases of the previous measurements \cite{Terashima_npj_2024}. 
From much larger $H_{\rm c2}$ than the expected Pauli-depairing field and the deviation mentioned above, it is considered that the robustness to $H$ would be the intrinsic characteristics in the Ta$_{1.6}$Te superconductor. 
%%%%%%%%%%%%%%%%%%%%%%%%%%%%%%%%%%%%%%%%%%%%%
\begin{figure}
    \centering
    \includegraphics[width=\linewidth]{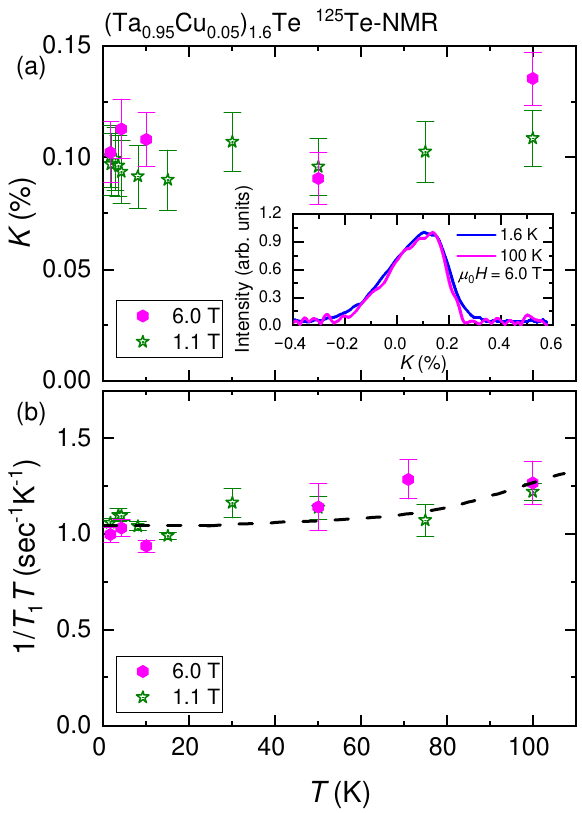}
    \caption{(a) Temperature dependence of the Knight shift of \ce{(Ta_{0.95}Cu_{0.05})_{1.6}Te} at \SI{6.0}{T} and \SI{1.1}{T} in the normal state. 
    The inset shows typical NMR spectrum in normal state at \SI{1.6}{K} and \SI{100}{K} at \SI{6.0}{T}. 
    (b) \ce{^{125}Te} nuclear spin-lattice relaxation rate $1/T_{1}T$ of \ce{(Ta_{0.95}Cu_{0.05})_{1.6}Te} at \SI{6.0}{T} and \SI{1.1}{T} in the normal state. 
    The dashed line is a guide to the eye.}
    \label{normal}
\end{figure}
%%%%%%%%%%%%%%%%%%%%%%%%%%%%%%%%%%%%%%%%%%%%%%%

Next, we refer to the normal-state magnetic properties. 
The normal-state \ce{^{125}Te} NMR spectrum observed at $\mu_0 H \sim 6$ T is shown in the inset of Fig.~\ref{normal} (a).
The \ce{^{125}Te} NMR spectrum shows a weak anisotropic shape, which is $T$ independent up to \SI{100}{K}.
%The anisotropy in the spectrum becomes less pronounced with decreasing $H$ as shown later, suggesting that the Knight-shift anisotropy is the origin of the anisotropic shape. 
Figure.~\ref{normal} (a) shows the $T$ dependence of the Knight shift measured at $\sim$ \SI{6}{T} and $\sim$ \SI{1.1}{T}. 
The Knight shift ($\sim$ \SI{0.1}{\%}) is almost $H$ and $T$ independent from \SI{1.6}{K} to \SI{80}{K}, consistent with previous observations in other QCs \cite{Jeglic_PRB_2005,Gavilano_PRL_1997,Yasuoka_JPSJ_1986}. 

The $T$ dependence of $1/T_{1}T$ in the normal state of $\sim$ \SI{6}{T} and $\sim$ \SI{1.1}{T} is shown in Fig.~\ref{normal} (b). 
As shown with the eye-guide (dashed) line, $1/T_{1}T$ is nearly $T$ independent below \SI{20}{K}, and starts to increase gradually above 20 K. 
It is considered that the low-$T$ plateau indicates that conduction-electron (Korringa) relaxation process is dominant as in conventional metals. 
A slight upturn of $1/T_{1}T$ at higher $T$ has also been reported in QCs \cite{Hill_PRB_1994,Gavilano_PRL_1997,Dolinsek_PRB_2001,Aphi_PRB_1999,Tang_PRL_1997,Dloinsek_PRB_2000,Jeglic_PRB_2005}.
Generally, in weakly correlated metals, the spin part of the Knight shift $K$ and $1/T_{1}T$ are related with the Korringa relation \cite{Korringa_Physica_1950}.
\begin{comment}
\begin{align}
    K \;=\; \sqrt\frac{\hbar}{4\pi k_{\mathrm{B}}\,T_{1}T}\Bigl(\frac{\gamma_{e}}{\gamma_{n}}\Bigr),
    \label{Korringa}
\end{align}
where $\gamma_{e}$ and $\gamma_{n}$ are the electronic and nuclear gyromagnetic ratios, respectively. 
Equation \eqref{Korringa} allows us to estimate spin part of $K$ from a measured $1/T_1$.      
\end{comment}
The measured $1/T_{1}T = 1.04\ \mathrm{s^{-1}K^{-1}}$ gives $K =$ \SI{0.16}{\%} which is $\sim 1.6$ times larger than the observed $K$.
%from eq. \ref{Korringa}. 
It is reasonable to consider that the observed $K$ arises from the spin-part Knight shift and that the orbital shift $K_{\rm orb}$ would be almost zero. 
The similar Korringa relation was also reported in the d–AlNiCo QC \cite{Jeglic_PRB_2005}.
%According to the hierarchically variable-range hopping model, the increase in $1/T_1T$ at sufficiently high temperatures originates from electron hopping assisted by acoustic phonon, which causes the conduction electron density to grow \cite{Dolinsek_PRB_2001}. This dynamic behavior is not captured by the Knight shift \(K\), since the latter reflects only the static density of states at the Fermi level.

%%%%%%%%%%%%%%%%%%%%%%%%%%%%%%%%%%%%%%%%
\begin{figure}
    \centering
    \includegraphics[width=\linewidth]{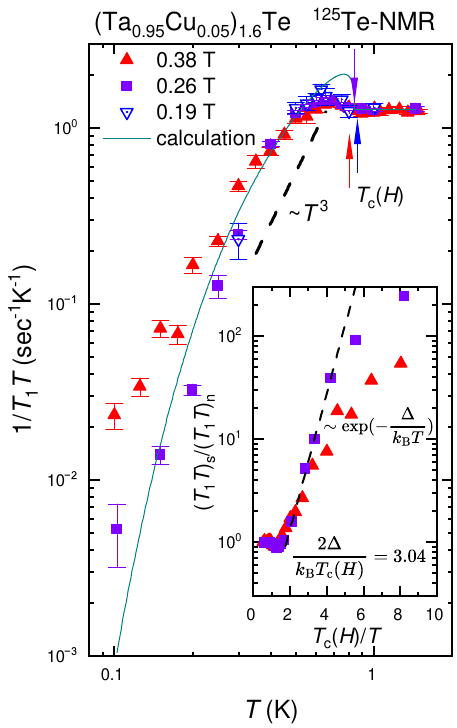}
    \caption{\ce{^{125}Te} nuclear spin-lattice relaxation rate divided by $T$ $1/T_{1}T$ of \ce{(Ta_{0.95}Cu_{0.05})_{1.6}Te} at \SI{0.38}{T}, \SI{0.26}{T}, and \SI{0.19}{T} below \SI{1.5}{K}. 
    The solid curve is the calculation for an conventional $s$-wave full-gap model with $2\Delta(0)/k_{\mathrm{B}}T_{\mathrm{c}} = 3.04$ and $\delta/\Delta(0) = 0.4$. 
    The Inset shows the plot of $(T_{1}T)_{\mathrm{s}}/(T_{1}T)_{\mathrm{n}}$ versus $T_{\mathrm{c}}(H)/T$ at \SI{0.26}{T} and \SI{0.38}{T}. 
    The linear dependence (dashed line) indicates an exponential decay of $1/T_{1}T$.}
    \label{T1_dilution}
\end{figure}
%%%%%%%%%%%%%%%%%%%%%%%%%%%%%%%%%%%%%%%%%%%%%%%%%
The $1/T_{1}T$ measured with the dilution refrigerator under several magnetic fields are plotted in Fig.~\ref{T1_dilution}.
In the SC state, a clear Hebel–Slichter (HS) peak just below $T_{\mathrm{c}}$ and an exponential decrease at lower $T$ shown in Fig.~\ref{T1_dilution} were observed, which are a hallmark of $s$-wave superconductivity with a nodeless full gap. 
The HS-peak amplitude becomes slightly larger at \SI{0.19}{T} than at \SI{0.38}{T}.
This indicates that the HS peak is suppressed with applied magnetic fields.
In addition, the application of larger magnetic fields also makes $1/T_1T$ increase at low temperatures.
This effect is understood as the relaxation arising from vortex cores, where quasiparticles are recovered to the normal-state by the applied fields.

The inset of Fig.~\ref{T1_dilution} shows the plot of $(T_1T)_{\rm s}/(T_1T)_{\rm n}$ as a function of $T_{\mathrm{c}}(H)/T$, where $1/(T_{1}T)_{\mathrm{s}}$ and $1/(T_{1}T)_{\mathrm{n}}$ are $1/T_1T$'s in the SC and normal states, respectively.     
From the Arrhenius plot of the result measured at \SI{0.26}{T}, the SC gap $2\Delta/k_{\mathrm{B}}T_{\mathrm{c}}$ in \ce{(Ta_{0.95}Cu_{0.05})_{1.6}Te} is estimated to be 3.04, which is slightly smaller than the BCS value (3.52).   
The smaller $\Delta$ indicates weak coupling superconductivity, consistent with the specific-heat measurement \cite{Tokumoto_Natcommun_2024}.

The observed $1/T_{1}T$ behavior is compared with a $s$-wave model calculation, which consistently explains the behavior in conventional superconductors.     
In Fig.~\ref{T1_dilution}, the solid curve is the $s$-wave model with full gap, which fits well below $T \lesssim 0.5\,T_{\mathrm{c}}$. 
Here, we used a rectangular broadening of the quasiparticle density of states (DOS), whose width is $2\delta$ and height is $1/2\delta$. 
In the model, $1/T_1T$ and the broadened DOS are expressed as follows \cite{Hebel_PR_1959}:
\begin{align*}
\frac{(T_{1}T)_{\mathrm{n}}}{(T_{1}T)_{\mathrm{s}}}
&=\frac{2}{k_{\mathrm{B}}T}\int_{0}^{\infty}N_{\mathrm{s}}(E)^2
\biggl(1+\frac{\Delta^2}{E^2}\biggr)f(E)\bigl[1-f(E)\bigr]\,dE,\\
N_{\mathrm{s}}(E)
&=\frac{1}{2\delta}\int_{E-\delta}^{E+\delta}\frac{E'}{\sqrt{E'^2-\Delta^2}}\,dE'.
%\theta\bigl(|E'|-\Delta\bigr)
\end{align*}
We used $2\Delta(0)/k_{\mathrm{B}}T_{\mathrm{c}}=3.04$, which was evaluated from above Arrhenius plot, and $\delta/\Delta(0)=0.4$. 
Although the large broadening of DOS is introduced, the observed suppression of the HS peak just below $T_c$ cannot be explained.
We also tentatively fit the experimental $1/T_1T$ data to a full-gap model without the coherence factor, which corresponds to the triplet $s$-wave pairing \cite{Cao_PRL_2020}.
We can fit the experimental results as in the case with the coherence factor, but $1/T_1T$ just below $T_c$ cannot be fitted consistently, as discussed in the supplemental materials \cite{SM}.  
It is noted that the HS peak becomes larger in ``dirty'' superconductors, because the electron scattering mixes Bloch states, and thus the anisotropy in the SC gap is essentially eliminated in very dirty superconductors with the electron mean free paths smaller than SC coherence length \cite{DEMacLaughlin1976}.  
The fitting results strongly suggest that there would be some effects for the suppression of the HS peak.

\begin{figure}
    \centering
    \includegraphics[width=\linewidth]{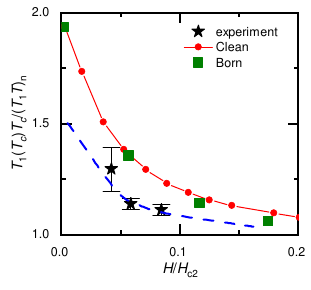}
    \caption{Height of the HS peak as a function of $H/H_{\mathrm{c2}}$. 
    Star symbols show the present result on \ce{(Ta_{0.95}Cu_{0.05})_{1.6}Te}. 
    The solid line (clean limit) and square symbols (Born limit) are theoretical results from the Eilenberger calculations \cite{Tanaka_PRB_2015}.}
    \label{Theory_comp}
\end{figure}

An effect we considered is the magnetic field. 
%Although the magnetic-field application is inevitable for the $1/T_1$ measurement with $^{125}$Te NMR because of $I$ of $^{125}$Te being 1/2, 
It is well known that the Volovik effect works to suppress the HS peak in conventional $s$-wave superconductors \cite{volovik_JETP_1993}.
This was studied theoretically and experimentally \cite{Ding_PRB_2016,Tanaka_PRB_2015}. 
Figure \ref{Theory_comp} shows the height of the HS peak observed in \ce{(Ta_{0.95}Cu_{0.05})_{1.6}Te} as a function of $H/H_{\mathrm{c2}}$.
In the figure, the Eilenberger-theory results in the Born limit and the clean-limit calculations are also plotted \cite{Tanaka_PRB_2015}. 
Although the HS-peak height measured in a conventional-superconductor CaPd$_2$As$_2$ quite agrees with these curves \cite{Ding_PRB_2016}, our QC data (stars) are systematically below both theoretical curves.
We consider that the strong suppression of the HS peak arises from the DOS-edge smearing caused by quasiperiodicity, which was recently suggested in Penrose-lattice calculations \cite{Takemori_PRB_2020}. 
Hence, the unusually strong HS-peak suppression (almost absence of a sharp Bogoliubov peak) in \ce{(Ta_{0.95}Cu_{0.05})_{1.6}Te} would be intrinsic properties related to the quasiperiodic structure. 

%%%%%%%%%%%%%%%%%%%%%%%%%%%%%%%%%%%%%%%%%%%%%%%%%%%
%%%%%%%%%%%%%%%%%%%%%%%%%%%%%%%%%%%%%%%%%%%%%%%%%%%

\begin{figure}
    \centering
    \includegraphics[width=\linewidth]{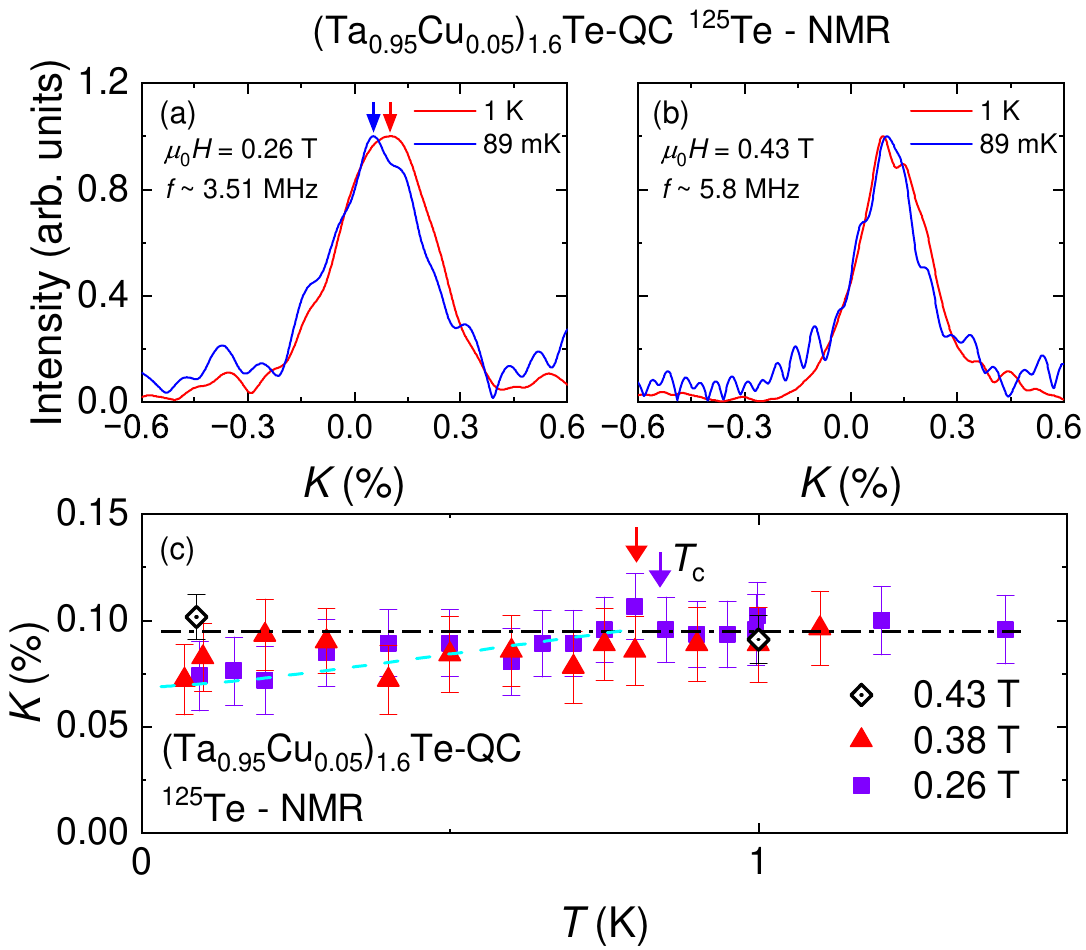}
    \caption{Typical \ce{^{125}Te} NMR spectra in the normal (a) and SC (b) states under \SI{0.26}{T} and \SI{0.43}{T}, respectively. 
    Arrows indicate the spectral peak. (c) Temperature dependence of the Knight shift measured at various fields. The dot-dashed and dashed lines are guides to the eye.}
    \label{shifts}
\end{figure}

Next, we discuss the NMR-spectrum variation in the SC state. 
Figures. \ref{shifts} (a) and (b) show \ce{^{125}Te}-NMR spectrum measured at 1 K ( $ > T_{\mathrm{c}}$) and 89 mK ( $ < T_{\mathrm{c}}$) .
Sharp NMR spectra with a linewidth of $\sim\SI{20}{kHz}$ was observed in such a small field in \ce{(Ta_{0.95}Cu_{0.05})_{1.6}Te}. 
A similar sharp NMR spectrum was reported in \ce{^{31}P}-NMR of amorphous \ce{(Mo_{0.5}Ru_{0.5})_{80}P_{20}} with $\sim\SI{15}{kHz}$ linewidth \cite{AliagaGuerra_solidstatecommun_1979}.
Although the sharp NMR spectrum seems to be counterintuitive to the lack of translational symmetry, the homogeneous electronic state is realized in these compounds without the translational symmetry.
%This is because the nuclei with $I=1/2$ carry no quadrupole moment and do not couple to the electric field gradient (EFG), which broadens NMR spectrum because of inhomogeneity. In fact, the broad spectra ($\sim\SI{100}{kHz}$) are observed in QC NMR studies using nuclei with $I > 3/2$ \cite{Jeglic_PRB_2005,Shastri_PRB_1994,Shastri_PRB_1995}.

It was found that the NMR line shape is almost unchanged when entering the SC state.
The heat-up test just after the NMR pulses and the sharp decrease of NMR intensity just below $T_{\mathrm{c}}$ confirm that the NMR measurements were done in the SC state, as discussed in supplemental material \cite{SM}. 
In general, in clean $s$-wave type-II superconductors, vortices create characteristic inhomogeneous internal fields that broaden the NMR line in an asymmetric fashion \cite{Tanaka_PRB_2015}, which is called ``Redfield pattern'' \cite{Redfield_PR_1967}.
%It was quite unusual that the NMR linewidth does not show such a asymmetric broadening below $T_{\mathrm{c}}$, although ac susceptibility exhibits a strong diamagnetic signal. 
The absence of the linewidth broadening indicates a relatively homogeneous local field at the \ce{Te} sites, implying a large magnetic penetration depth and weak diamagnetism, although the ac susceptibility exhibits a robust diamagnetic shielding, which indicates strong diamagnetism. 

This inconsistency might be explained by intrinsic vortex pinning in QCs \cite{Nagai_PRB_2022}. 
In a QC superconductor, atomic‐scale variation among the order parameters pins magnetic vortices at sites where the order parameter is locally suppressed. 
As these sites induce only weak SC currents, the local diamagnetic screening around each vortex would be small.
Consequently, the overall field distribution inside the superconductor remains relatively uniform, preventing the Redfield pattern in the SC NMR spectrum. 
In contrast, the ac susceptibility, being a bulk probe, is mainly dominant at the surface region, so it still exhibits a pronounced diamagnetic signal.

Finally, we discuss the $T$ dependence of the Knight shift in the SC state.
Figure \ref{shifts} (c) displays the $T$ dependence of the Knight shift below 1.4 K, measured at several magnetic fields. 
Below $T_{\mathrm{c}}$, the Knight shift decreases by only $\sim\SI{0.03}{\%}$ at \SI{0.26}{T}, and this decrease is almost suppressed at \SI{0.43}{T}. 
In general, the Knight shift in the SC state can be decomposed as follows,
\begin{align*}
K(T,H) = K_{\mathrm{spin}}(T,H) + K_{\mathrm{orb}} + K_{\mathrm{dia}}(T,H).
\end{align*}
Here, $K_{\mathrm{spin}}$ is the spin component of the Knight shift $K_{\mathrm{spin}}(T,H)$, which is proportional to the spin susceptibility with a hyperfine coupling constant, $K_{\mathrm{orb}}$ is the $T$-independent Van-Vleck orbital contribution, and $K_{\mathrm{dia}}$ is the SC diamagnetic shift, which would be negligibly small due to the almost absence of NMR linewidth broadening.
In fact extremely small $K_{\mathrm{dia}}$, estimated with theoretical calculation and experimental values, is in supplemental material \cite{SM}. 
It is well known that $K_{\mathrm{spin}}(T,H)$ decreases following the Yosida function in a conventional $s$-wave superconductor \cite{Fine_PL_1969,Yoshida_PR_1958}.
As discussed above, it is considered that the observed $K$ arises from the spin part and the orbital shift $K_{\mathrm{orb}}$ would be almost zero. 
\begin{comment}
    In addition, $K_{\mathrm{dia}}$ is approximately calculated by using the following theoretical expression\cite{Brandt_PRB_2003}.
\begin{align}
    K_{\mathrm{dia}}= -(1-N)\frac{H_{\mathrm{c2}}}{H}\frac{\ln{(\frac{H_{\mathrm{c2}}}{H}})}{4\kappa^2}\times 100 (\%)
\end{align}
where $\mu_{0}H_{\mathrm{c2}}$, $N$( = 1/3) and $\kappa$ are the upper critical field, the demagnetization factor, and the Ginzburg-Landau parameter, respectively.
From the specific-heat measurement \cite{Tokumoto_Natcommun_2024}, the SC critical field $H_{\mathrm{c}}(0)$ was evaluated as \SI{26.5}{Oe}. 
Using the experimental values of $\mu_{0} H_{\mathrm{c2}}(0)\sim$ \SI{4.6}{T} and $\kappa \approx 1.2 \times 10^3$, the $K_{\mathrm{dia}}$ is estimated as $-5.9\times10^{-4} (\%)$, which is negligibly small.
This is consistent with the almost absence of NMR linewidth broadening. 
\end{comment}
Therefore, the Knight shift decrease mostly originates from $K_{\rm spin}$.
%As shown in Fig.~\ref{shifts}(c), the Knight shift measured in \ce{(Ta_{0.95}Cu_{0.05})_{1.6}Te} decreases only $\sim 0.03\%$ below $T_{\mathrm{c}}$. 

Such a small Knight-shift decrease is often attributed to spin–orbit–mediated scattering in dirty metals \cite{Wright_PRL_1967,Hines_PRB_1971,Hines_PRL_1967,Miyake_PRB_2000,Yamada_2025_PRB}. 
In this case, spin flips mix plane-wave states of different spin states, so they no longer form energy eigenstates \cite{Anderson_PRL_1959}. 
In contrast, as the wave vector cannot be well-defined in QC because of the absence of translational symmetry, new coherent state would be formed as in the case of the dirty metals.    
Although both scenarios suppress the Knight-shift decrease, the former relies on spin–orbit scattering in a dirty metal, but the latter stems directly from the quasiperiodic lattice structure. 
Thus, we suggest that the reduced Knight-shift decrease observed in \ce{(Ta_{0.95}Cu_{0.05})_{1.6}Te} would be a characteristic feature of QC superconductors.
The unchanged Knight-shift in the SC state above \SI{0.4}{T} seems to be consistent with the absence of the Pauli effects near $H_{\rm c2}$.     

We briefly refer to the parity mixing \cite{Sigrist_AIP_Conf_Proc_2009} and the possibility of the spin-triplet superconductivity \cite{Cao_PRL_2020}. 
Since the approximant crystals (\ce{Ta97Te60} and \ce{Ta181Te112}) have no mirror symmetry along the $c$ axis \cite{Wolfgang_doctorial_thesis, Cain_PNAS_2020, Conrad_CEJ_2002}, there would be no mirror symmetry along the $c$ axis in QC \ce{Ta_{1.6}Te}. 
In addition, the theoretical study suggests that the spin-triplet pairing is possible even with the $s$-wave pairing on QCs, because the spin state and angular momentum in a SC paring are uncorrelated due to lack of translational symmetry on the QC lattice \cite{Cao_PRL_2020}. 
Therefore, the marginal decrease of the Knight shift in low field and the much larger $H_{\mathrm{c2}}$ than the Pauli-limit field might suggest the parity mixing \cite{Sigrist_AIP_Conf_Proc_2009}, and the predominance of the triplet component even in the low-$H$ far below $H_{\rm c2}$.
%In addition, the Knight shift probes the local magnetic susceptibility only in the immediate vicinity of each nucleus, it predominantly reflects those regions of weak diamagnetism where vortices are pinned.  
%%%%%%%%%%%%%%%%%%%%%%%%%%%%%%%%%%%%%%%%%
%%%%%%%%%%%%%%%%%%%%%%%%%%%%%%%%%%%%%%%%%%%%

%\section{Conclusion}
In conclusion, our $^{125}$Te-NMR measurements on \ce{(Ta_{0.95}Cu_{0.05})_{1.6}Te} revealed that the SC gap has an isotropic $s$-wave character with $2\Delta/k_BT_{\rm c} \sim 3.04$. 
The smaller HS peak just below $T_{\mathrm{c}}$ seems to be consistent with the extremely weak divergence at $\Delta$ in the quasiparticle DOS, which was theoretically suggested in the SC character in QCs.   
The almost absence of the broadening and marginal shift in the NMR spectrum below $T_c$ in small $H$ as well as the absence of the Pauli-depairing effect and the continuous increase of $H_{\mathrm {c2}}$ down to $T = 0$ are quite unusual and unambiguously strange. 
These behaviors can be seemingly understood with the scenario of the dirty-metal state arising from the QC structure, but might suggest the occurrence of the unusual SC state such as FFLO state \cite{Sakai_PRR_2019} or parity mixing SC state \cite{Sigrist_AIP_Conf_Proc_2009,Cao_PRL_2020}.
$^{125}$Te-NMR measurement on single-crystal samples under high $H$ is crucially important to clarify the SC state in QC Ta$_{1.6}$Te.  

%\section{Acknowledgement}
The authors would like to thank S. Watanabe, K. Imura and N. Kabeya for fruitful discussions. 
This work was supported by Grants-in-Aid for Scientific Research (KAKENHI Grant No. JP20KK0061, No. JP20H00130, No. JP21K18600, No. JP22H04933, No. JP22H01168, No. JP23H01124, No. JP23K19022, No. JP23K22439, No. JP23K25821, No. 24K00590, No. 25KJ1659, and No. JP23K04355 ) from the Japan Society for the Promotion of Science, by JST SPRING(Grant No. JPMJSP2110) from Japan Science and Technology Agency, by research support funding from The Kyoto University Foundation, by ISHIZUE 2024 of Kyoto University Research Development Program, by Murata Science and Education Foundation, by the JGC-S Scholarship Foundation, Iketani Science and Technology Foundation , and by JST-CREST program (Grant No. JPMJCR22O3; Japan).
Work at IMR Tohoku University was supported under the IMR-GIMRT Program (Proposal Number 202212-HMKPB-0007, 202212-HMKPB-0008, 202212-IRKAC-0041, 202312-HMKPB-0019, 202312-HMKPB-0022, 202312-IRKAC-0007).
In addition, liquid helium is supplied from the Low Temperature and Materials Sciences Division, Agency for Health, Safety and Environment, Kyoto University. 
\bibliographystyle{apsrev4-1}
\bibliography{Ta1.6Te}

\clearpage

\section*{Supplemental material}
\renewcommand{\thesubsection}{S\arabic{subsection}}
\setcounter{subsection}{0}

\renewcommand{\thefigure}
%{S\arabic{subsection}.\arabic{figure}}
{S\arabic{figure}}
\setcounter{figure}{0}

\subsection{Relaxation curves}
\begin{figure}[htbp]
    \centering
    \includegraphics[width=\linewidth]{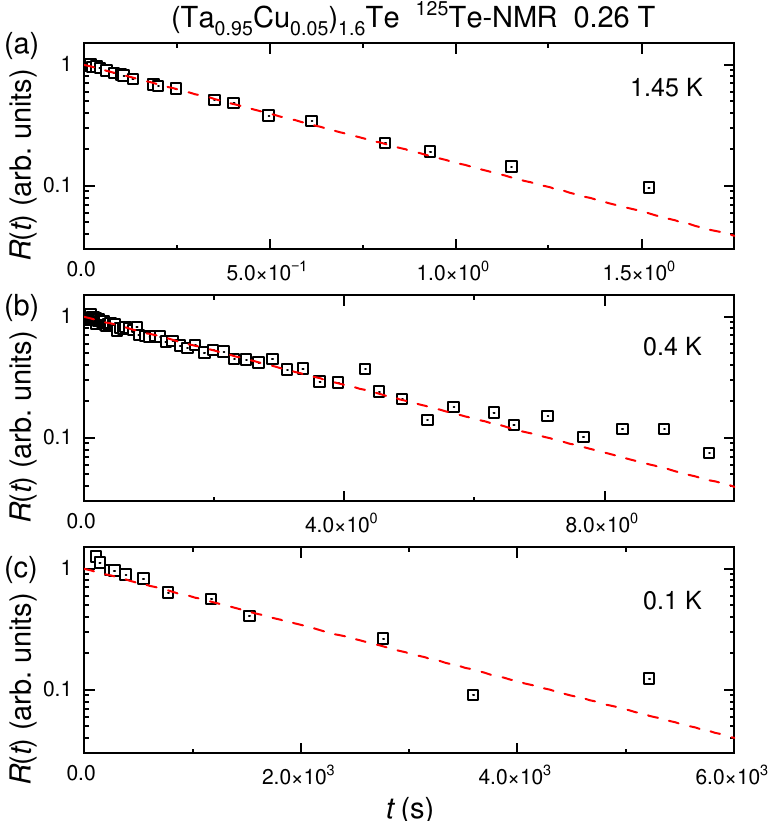}
    \caption{\ce{^{125}Te} relaxation curves $R(t)$ at \SI{1.45}{K} (a), normal state, \SI{0.4}{K} (b), SC state, and \SI{0.1}{K} (c), SC state under \SI{0.26}{T}. The red dashed lines are fits to single-exponential decay relaxation curves $R(t) \equiv 1 - M(t)/M(\infty)$, with the nuclear magnetization $M(t)$ at $t$ after the saturation pulse, at \SI{1.45}{K} (normal state), \SI{0.4}{K} (SC state), and \SI{0.1}{K} (SC state) under \SI{0.26}{T} respectively, }
    \label{relaxation}
\end{figure}

%\label{sec:relaxation curve}
Here, we discuss how to determine $1/T_1$ value from the measurements.
$1/T_{1}$ was evaluated by fitting the relaxation curve of the nuclear magnetization after its saturation to a single component $R(t) \propto \exp(-t/T_1)$, which is a theoretical function for $I$ = 1/2.
Here, relaxation curves $R(t)$ is expressed as $R(t) \equiv 1 - M(t)/M(\infty)$ with the nuclear magnetization $M(t)$ at $t$ after the saturation pulse. 
As shown in Figs.~\ref{relaxation} (a) and (b), $1/T_1$ was determined with the single-component function in the normal state and the SC state above 0.3 K.
However, a slightly faster relaxation component emerges below \SI{0.3}{K}, as seen in Fig.~\ref{relaxation} (c).
This is probably due to the instantaneous heating effect just after NMR Radio Frequency (RF) pulses, thus the slow component, which can be fit with a single component $R(t)$, was adopted as $1/T_{1}$.

%%%%%%%%%%%%%%%%%%%%%%%%%%%%%%%%%%%%%%%%%
%%%%%%%%%%%%%%%%%%%%%%%%%%%%%%%%%%%%%%%%%%%%%%%%%%

\subsection{Heating effect by NMR RF-pulses}
%%%%%%%%%%%%%%%%%%%%%%%%%%%%%%%%%%%%%%%%%
\begin{figure*}[htbp]
    \centering
    \includegraphics[width=\linewidth]{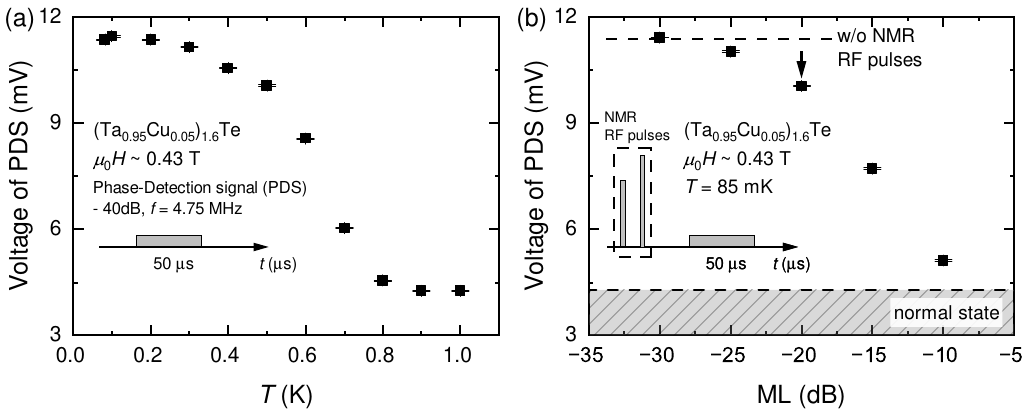}
    \caption{(a) Temperature dependence of the voltage of the imaginary part for the output signal from the NMR receiver after a phase-detection signal (PDS) was inputted. The pulse width and the output power of the PDS is 50 $\mu$s. (b) The dependence of the master level (ML) applied before the PDS at \SI{0.43}{T}. The dependence was measured in the \ce{^{3}He}/\ce{^{4}He} mixture temperature of \SI{85}{mK}. The pulse sequence of the test is shown in the inset of the Figure. The voltage shown by "w/o NMR RF pulses" is the voltage of the PDS without NMR RF pulses, and the voltage shown by "Normal state" is the voltage of PDS expected in the normal state. In the actual NMR measurement, we applied RF pulses at \SI{-20}{dB} (black arrow). }
    \label{heatup_check}
\end{figure*}
%%%%%%%%%%%%%%%%%%%%%%%%%%%%%%%%%%%%%%%%%
 The superconductivity just after the RF pulses for the NMR-signal observation was confirmed by the following heat-up test using the same set-up as the NMR measurement with dilution refrigerator.
 Recently, such test has been employed in NMR measurements on various superconductors these days \cite{Pustogow2019,Ishida_JPSJ_2020, Fujibayashi_JPSJ_2022,Kinjo_Sciadv_2023}.
 We applied a weak RF pulse with the pulse width of 50 $\mu$s and the frequency of which is \SI{4.75}{MHz}. 
 We measured the voltage of the imaginary part of the output signal from the NMR receiver. This weak RF pulse is called "phase-detection signal (PDS)". The strength of the PDS is \SI{-40}{dB}.

First, we measured the $T$ dependence of the voltage of the imaginary part for the output signal of the PDS.
In the measurement, we applied $\mu_{0}H\sim$ \SI{0.43}{T} as in the present NMR measurement.
Figure. \ref{heatup_check} (a) shows the $T$ dependence of the voltage.
The voltage changed at $T_{\mathrm{c}}$ due to the change of the impedance of the NMR tank circuit, because the inductance of the NMR coil with the sample was changed when superconductivity sets in.

Next, we applied two RF pulses corresponding to $\pi/2$ and $\pi$ pulses just before the PDS as in the NMR measurements.
The pulse width of each pulse is fixed 30 $\mu$s and the voltage of the $\pi/2$ pulse is half of the $\pi$ pulse.
The pulse sequence of the test is shown in the inset of Fig. \ref{heatup_check} (b).
We measured the dependence of the voltage of the PDS on the master level (ML) intensity of two RF pulses in the \ce{^{3}He/^{4}He} mixture temperature of \SI{85}{mK}.
Figure. \ref{heatup_check} (b) shows that the voltage of the PDS is unchanged up to \SI{-25}{dB}, and from the comparison between Fig. \ref{heatup_check} (a) and (b), we conclude that superconductivity is maintained up to ML $\sim$ \SI{-10}{dB}.

In our actual NMR spectra measurement at \SI{89}{mK}, we applied RF pulses at \SI{-20}{dB}.
Therefore, we can say that the electronic state remains in the SC state, although the heating effect by the RF field is not completely negligible.

%%%%%%%%%%%%%%%%%%%%%%%%%%%%%%%%%%%%%%%%%
\begin{figure}[htbp]
    \centering
    \includegraphics[width=\linewidth]{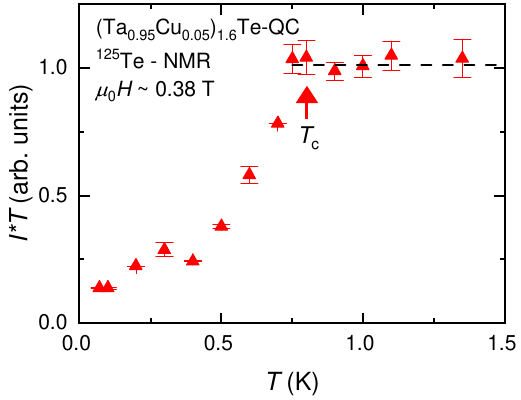}
    \caption{Temperature dependence of the product of NMR signal intensity $I$ and temperature $T$ (i.e. $I\times T$), normalized to its value just above $T_{\mathrm{c}}$.}
    \label{IT}
\end{figure}
%%%%%%%%%%%%%%%%%%%%%%%%%%%%%%%%%%%%%%%%%%%%%%%%%%
In addition, Fig. \ref{IT} shows $T$ dependence of the product of NMR signal intensity $I$ and temperature $T$ (i.e. $I\times T$).
While $I\times T$ is almost constant in the normal state, the $I\times T$ decreases substantially below $T_c$ because of the SC diamagnetic effect.  
Therefore, we can say that we observed the NMR spectra in the SC state.
Thus, we conclude that almost no shift and no broadening of the NMR spectrum is the intrinsic SC properties in \ce{(Ta_{0.95}Cu_{0.05})_{1.6}Te}.

\newpage

\subsection{Numerical estimation of $K_{\mathrm{dia}}$}
Here, we estimate the shift by the SC diamagnetism ($K_{\mathrm{dia}}$) with the theoretical equation and experimental values. 
$K_{\mathrm{dia}}$ is approximately calculated by using the following theoretical expression \cite{Brandt_PRB_2003}.
\begin{align}
    K_{\mathrm{dia}}= -(1-N)\frac{H_{\mathrm{c2}}}{H}\frac{\ln{(\frac{H_{\mathrm{c2}}}{H}})}{4\kappa^2}\times 100  (\%)
\end{align}
where $\mu_{0}H_{\mathrm{c2}}$, $N$( = 1/3) and $\kappa$ are the upper critical field, the demagnetization factor, and the Ginzburg-Landau parameter, respectively.
From the specific-heat measurement \cite{Tokumoto_Natcommun_2024}, the SC critical field $\mu_0 H_{\mathrm{c}}(0)$ was evaluated as \SI{2.65}{mT}. 
Using SC parameters summarized in Tab.\ref{tab:SC_parameter}, $K_{\mathrm{dia}}$ is estimated as $-5.9\times10^{-4}$ \%, which is negligibly small.
This is consistent with the almost absence of NMR linewidth broadening in the SC state. 
\begin{table}[]
\caption{SC parameters for \ce{(Ta_{0.95}Cu_{0.05})_{1.6}Te}}
    \label{tab:SC_parameter}
    \begin{tabular}{|c|c|}
    \hline
    $H_{\mathrm{c}2}$ & \SI{4.6}{T} \\
    \hline
    $H_{\mathrm{c}}$ & \SI{2.65}{mT}\\
    \hline
    $\lambda$ &  $1.04\times10^5$ \AA\\
    \hline
     $\xi$ &  84.6 \AA\\
     \hline
     $\kappa (=\lambda/\xi)$ & $1.2\times 10^3$\\ 
     \hline
\end{tabular}
    
\end{table}

\subsection{Numerical calculation of $1/T_1$ with SC models}
As described in the main text, we tried to fit the $1/T_1$ results to numerical calculations with the SC models. 
We tentatively calculated $1/T_1$ with the full-gap SC models without the coherence factor, which corresponds to the spin-triplet pairing.   
In this case, the SC state $1/(T_1T)_{\mathrm{s}}$ normalized with the normal-state $1/(T_1T)_{\mathrm{n}}$ is given by the following expression:
\begin{align*}
\frac{(T_1T)_{\rm n}}{(T_1T)_{\rm s}}
&=\frac{2}{k_{\mathrm{B}}T}\int_{0}^{\infty}N_{\mathrm{s}}(E)^2
f(E)\bigl[1-f(E)\bigr]\,dE,\\
N_{\mathrm{s}}(E)
&=\frac{1}{2\delta}\int_{E-\delta}^{E+\delta}\frac{E'}{\sqrt{E'^2-\Delta^2}}\,dE'.
\end{align*}

%%%%%%%%%%%%%%%%%%%%%%%%%%%%%%%%%%%%%%
\begin{figure}
   \centering
    \includegraphics[width=\linewidth]{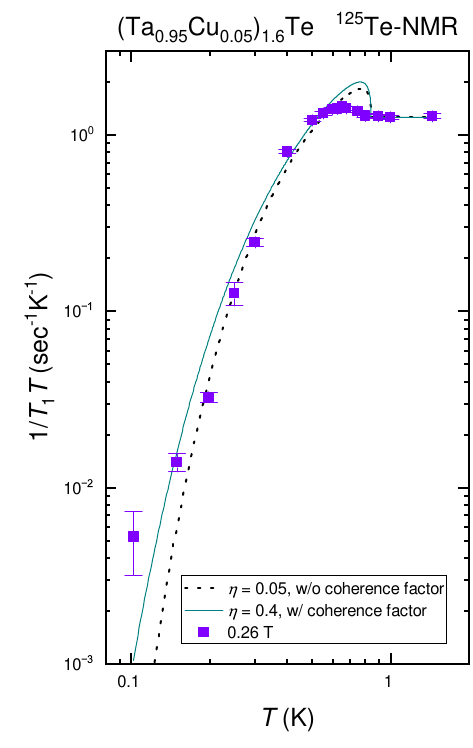}
    \caption{Numerical result for the nuclear spin-lattice relaxation rate $1/T_1$. The dotted line shows the calculation without the coherence factor ($2\Delta(0)/k_{\mathrm{B}}T_{\mathrm{c}}=3.04$, $\delta / \Delta(0) = 0.05$). The solid line shows the calculation with the coherence factor ($2\Delta(0)/k_{\mathrm{B}}T_{\mathrm{c}}=3.04$), identical to the solid curve in Fig.~\ref{T1_dilution}. The square markers represent the experimental data at \SI{0.26}{T}.}
    \label{nocoherence}
\end{figure}
%%%%%%%%%%%%%%%%%%%%%%%%%%%%%%%%%%%%%%
By setting $2\Delta(0)/k_{\mathrm{B}}T_{\mathrm{c}} = 3.04$, the same value as in the main text, experimental $1/T_1T$ data below $T_c / 2$ could be reasonably reproduced with the small broadening factor $\eta \equiv \delta/\Delta(0) = 0.05$, as shown in Fig.~\ref{nocoherence}.

%%%%%%%%%%%%%%%%%%%%%%%%%%%%%%%%
\begin{figure}
    \centering
    \includegraphics[width=\linewidth]{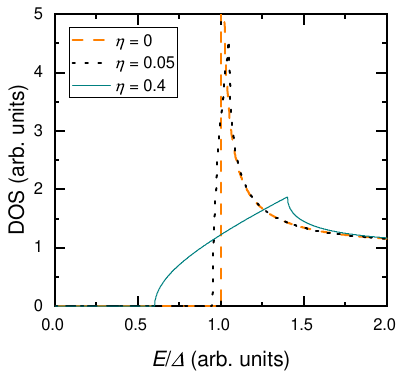}
    \caption{Broadened density of states (DOS). The dashed line ($\delta / \Delta(0) = 0$) shows the unbroadened DOS. The dotted line ($\delta / \Delta(0) = 0.05$) shows the broadened DOS used in the calculation without the coherence factor. The solid line ($\delta / \Delta(0) = 0.4$) shows the broadened DOS used in the calculation with the coherence factor.}
    \label{DOS_structure}
\end{figure}
%%%%%%%%%%%%%%%%%%%%%%%%%%%%%%%%%%%%%%%%
This result suggests that the possibility of spin-triplet pairing cannot be excluded from the $1/T_1T$ results. 
However, in order to reproduce the experimental behavior with the above SC model, the DOS edge becomes sharp, as illustrated in Fig.~\ref{DOS_structure}.

Such a sharp edge in the DOS is inconsistent with the theoretical predictions, which suggest that the divergence of DOS at $\Delta$ (Bogoliubov peak) should be smeared out due to a distribution of order parameters arising from the quasiperiodicity of the system~\cite{Takemori_PRB_2020}. 
Both calculations exhibit larger $1/T_1T$ near $T_{\mathrm{c}}$, which can be attributed to the divergence of the DOS near $T_{\mathrm{c}}$, even in the absence of the coherence factor. The fitting results by the $s$-wave model supports the theoretical suggestion that the divergence of DOS at $\Delta$ is significantly suppressed compared with the conventional case \cite{Takemori_PRB_2020}.

\end{document}